\documentclass[aps,prb,10pt,twocolumn,a4paper,superscriptaddress,nofootinbib,longbibliography,floatfix]{revtex4-2}

\usepackage{graphicx}
\usepackage{amsmath}
\usepackage{amssymb}
\usepackage{amsfonts}
\usepackage{color}
\usepackage{xcolor}
\usepackage[colorlinks = true,
            linkcolor = blue,
            urlcolor  = blue,
            citecolor = blue,
            anchorcolor = blue,
            unicode]{hyperref}

\begin{document}

\title{Equidistant resonance jumps in superconducting coplanar resonators driven by Abrikosov vortices}

\author{Dmitrii~S.~Kalashnikov}
\email{kalashnikov.ds@phystech.edu}
\affiliation{Moscow Institute of Physics and Technology, 141700 Dolgoprudny, Russia}

\author{Denis~Yu.~Vodolazov}
\affiliation{Moscow Institute of Physics and Technology, 141700 Dolgoprudny, Russia}
\affiliation{Institute for Physics of Microstructures, Russian
Academy of Sciences, 603950, Nizhny Novgorod, GSP-105, Russia}

\author{Ruslan~I.~Kinzibaev}
\affiliation{Moscow Institute of Physics and Technology, 141700 Dolgoprudny, Russia}
\affiliation{Dukhov Research Institute of Automatics (VNIIA), 127055 Moscow, Russia}

\author{Andrei~G.~Shishkin}
\affiliation{Moscow Institute of Physics and Technology, 141700 Dolgoprudny, Russia}
\affiliation{Dukhov Research Institute of Automatics (VNIIA), 127055 Moscow, Russia}

\author{Vasily~S.~Stolyarov}
\email{stolyarov.vs@phystech.edu}
\affiliation{Moscow Institute of Physics and Technology, 141700 Dolgoprudny, Russia}
\affiliation{Dukhov Research Institute of Automatics (VNIIA), 127055 Moscow, Russia}

\begin{abstract}
Superconducting coplanar resonators are key building blocks of cryogenic microwave circuits, yet their performance in perpendicular magnetic fields is ultimately limited by Abrikosov vortices. In this work we investigate the dependence of the transmission parameter $S_{21}$ of niobium quarter-wave coplanar resonators on perpendicular magnetic fields up to $40$ Oe and at temperatures between $18$ mK and $5$ K. Beyond the reversible Meissner regime, the entire resonance peak exhibits abrupt, staircase-like jumps as a function of magnetic field. Upon reversal of the field sweep, these jumps form an almost equidistant series with spacing $\Delta H \simeq 1.7$--$1.8$ Oe, which, in agreement with theoretical estimates, we interpret as signatures of multiple-vortex entry and exit events. Additionally, we observe the non-proportional responses of the resonant frequency and the internal quality factor that indicate a complex contribution of vortex and antivortex configurations. We expect that our results will stimulate further studies of large vortex–antivortex systems, explicitly accounting for their discrete nature.
\end{abstract}

 \maketitle

\section{Introduction}
Superconducting coplanar waveguides and resonators {underpin a broad range of cryogenic microwave} technologies, including superconducting quantum {processors}~\cite{Wallraff_2004}, {cryogenic} memory {elements} \cite{Kalashnikov_2024, Nulens_2025}, kinetic inductance detectors~\cite{Day_2003, Gao_2012}, and {general-purpose microwave component}s~\cite{Zmuidzinas_2012}. For these applications, significant efforts have been devoted to the development of high-quality-factor resonators~\cite{Noguchi_2019, Zikiy_2023, Poorgholam_2025}; however, their increased sensitivity to external perturbations makes the stability of their parameters an important issue.

The performance {of coplanar} resonators is strongly {affected by magnetic fields applied} perpendicular to the device plane. Such fields {can be unavoidable (residual fields, trapped flux in surrounding components) or intentionally applied to control on-chip elements such as} tunable qubits \cite{Chang_2023} or frequency-tunable resonators \cite{Palacios_2008}. {Understanding the magnetic-field response of planar resonators is therefore important both for mitigating loss and for enabling field-compatible circuit designs.}

At {sufficiently small} fields, a {coplanar resonator typically exhibits a weak and reversible parabolic shift of the resonant frequency, which reflects the field-induced increase of the kinetic inductance in the Meissner state}~\cite{Healey_2008}. {At higher fields,} Abrikosov vortices penetrate the superconducting film. {Each vortex carries a single flux quantum} $\Phi_0 \approx 2.068\times10^{-15}$~Wb {and introduces additional dissipation and inductance through the vortex core and the surrounding supercurrents \cite{Gittleman_1966, Coffey_Clem_1991}.} {As a result, vortices reduce} both the resonant frequency and the quality factor \cite{Bothner_2012}. Due to pinning, the magnetoresponse is strongly hysteretic and, for planar geometry, can often be described by
the Norris--Brandt--Indenbom (NBI) model~\cite{Norris_1970, Brandt_Indenbom_1993}. {A practical strategy for mitigating vortex-induced effects is to engineer flux screening and pinning landscapes, for example by perforating the film} (antidots)~\cite{Bothner_2017, Kroll_2019}. While a mixed vortex state is inherent to type-II superconductors like niobium, vortices can also nucleate in thin films of type-I materials, such as the commonly used aluminum \cite{Song_2009}.

Importantly, vortex penetration is frequently not a smooth process: the geometric/surface barrier for vortex entry can promote cooperative vortex avalanches. Extensive experimental and modeling efforts have been devoted to understanding this instability in diverse superconducting geometries — from bulk samples \cite{PhysRevB.74.054507_2006} and multifilamentary wires \cite{xue2024NatComm, renSUST_2025} to thin films \cite{PhysRevB.84.054537_2011, PhysRevB.94.054509_2016, Jiang_2020}

Notably, signatures of such avalanches have recently been observed in superconducting resonators \cite{Nulens_2023}. 
While vortex-resonator interaction is a primary source of decoherence and loss, they also provide a sensitive probe of nonequilibrium and collective vortex dynamics \cite{Nsanzineza_2014, Ghigo_2023}. Moreover, ongoing efforts to control Abrikosov vortices open the possibility of employing them as functional elements in superconducting logic circuits and memory cells \cite{Golod_2015, Keren_2023, Skog_2024}.

In this work, we perform high-resolution magnetospectroscopy from zero-field-cooled state of several coplanar resonators on two chips with different degree of superconducting film homogeneity. We demonstrate a reproducible staircase-like evolution of the resonance parameters and identify an almost equidistant sequence of jump fields that emerges immediately after reversing the field sweep direction. We show that the jump positions differ between resonators indicating a local origin of the phenomena, and strongly depend on the superconducting film quality.
To estimate the contribution of vortices to the resonator inductance, we employed a simple model derived from conventional vortex dynamics (Appendix \ref{Ap:model:vortex:Z}) and extend it with theoretical calculations obtained from numerical solutions of the time-dependent Ginzburg--Landau equation (Appendix \ref{Ap:numerical:vortex:Z}).

\begin{figure}[t!]
\begin{center}
\includegraphics[width=8.5cm]{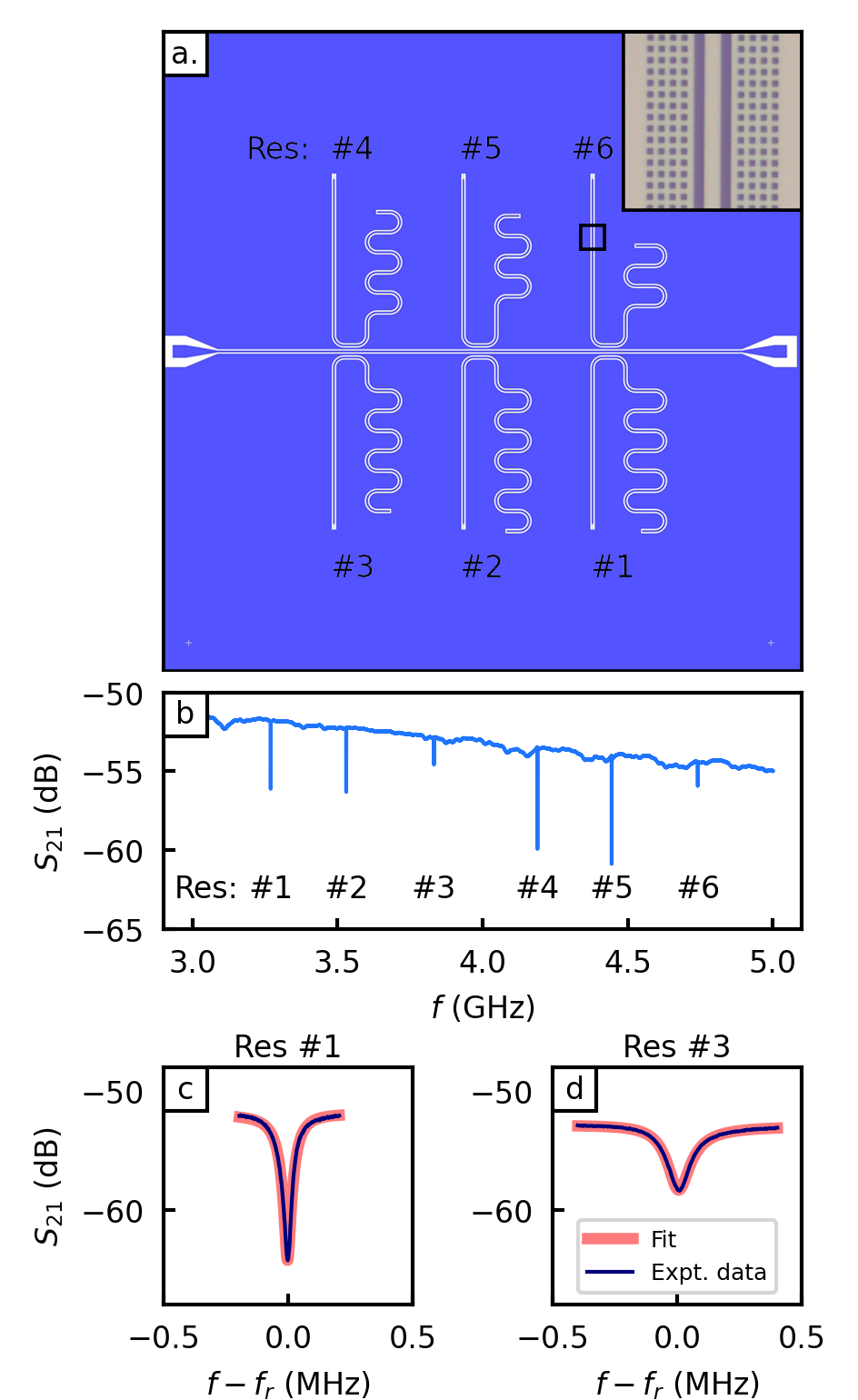}
\caption{Sample~A. (a) Design of the chip layout with six capacitively coupled (notch-type) quarter-wave coplanar resonators. The inset shows optical image of the resonator section. (b) Broadband transmission $S_{21}(f)$ showing six resonance dips. (c,d) Representative resonance dips for resonators \#1 and \#3 together with fits by an asymmetric Lorentzian line shape \cite{Probst_2015}.}
\label{fig: Design}
\end{center}
\end{figure}

\section{Experiment}

\subsection{Sample design and measurement protocol}

We investigate two chips (samples~A and~B) containing superconducting coplanar resonators patterned from a $100$-nm-thick niobium film. The fabrication details and the measurement setup are summarized in Appendix~\ref{appendix: Methods}. The layout of sample~A---the primary focus of this work---is shown in Fig.~\ref{fig: Design}(a). It contains six {quarter-wave} resonators of different lengths and thus different resonance frequencies. The resonators are capacitively coupled to a common feedline in a notch-type geometry. Additional data for sample~B are presented in Appendix~\ref{appendix: sample B}.

All resonators have the central line width $W = 30$~\textmu m and the gap to the ground $G = 17$~\textmu m, which sets the value of the wave impedance $Z_0 = 50~\Omega$. Both the {feedline and the resonators are surrounded by a perforated ground plane: a} square lattice of $10\times10$~\textmu m${}^2$ holes with a $10$~\textmu m {edge-to-edge separation (lattice period $20$~\textmu m). Such perforations are known to reduce detrimental vortex-induced effects by trapping magnetic flux \cite{Bothner_2017, Kroll_2019}.}

In the experiment, {we measure the complex transmission coefficient $S_{21}$ using a vector network analyzer. Each resonance appears as a sharp dip in $|S_{21}(f)|$ due to microwave power} absorption by the resonator. Figure~\ref{fig: Design}~(b) shows {a representative broadband frequency sweep over all six fundamental modes.}

We fit each resonance dip with an asymmetric Lorentzian line shape to extract the resonant frequency $f_r$ and the internal quality factor $Q_i$ \cite{Pozar_book, Probst_2015}. For a quarter-wave resonator the fundamental mode frequency and internal quality factor can be written as
\begin{equation}
\label{eq: definition_fr_Qi}
    f_r = \frac{1}{4l\sqrt{LC}}, \qquad
    Q_i = \frac{\pi}{2 R l}\sqrt{\frac{L}{C}},
\end{equation}
where $L$, $C$ and $R$ {are the per-unit-length inductance, capacitance and resistance, respectively,} $l$ is the length of a resonator.

The resonance dips {analyzed below are shown in Fig.~\ref{fig: Design}(c) and (d). At the base temperature} of $18$~mK and in absence of external magnetic field, we obtain for resonator~\#1: $f_r = 3.270$~GHz, $Q_i = 1.6\times10^5$; and for resonator~\#3: $f_r = 3.832$~GHz, $Q_i = 4.7\times10^4$.
The coupling quality factor $Q_c$, determined by the geometry of the coupling region between the resonator and the central feedline, is approximately $5\times10^4$ for all resonators on the chip. 
The corresponding lengths of resonators \#1 and \#3 equal 9.34 mm and 7.95 mm, respectively.

\begin{figure*}[ht!]
\begin{center}
\includegraphics[width=16cm]{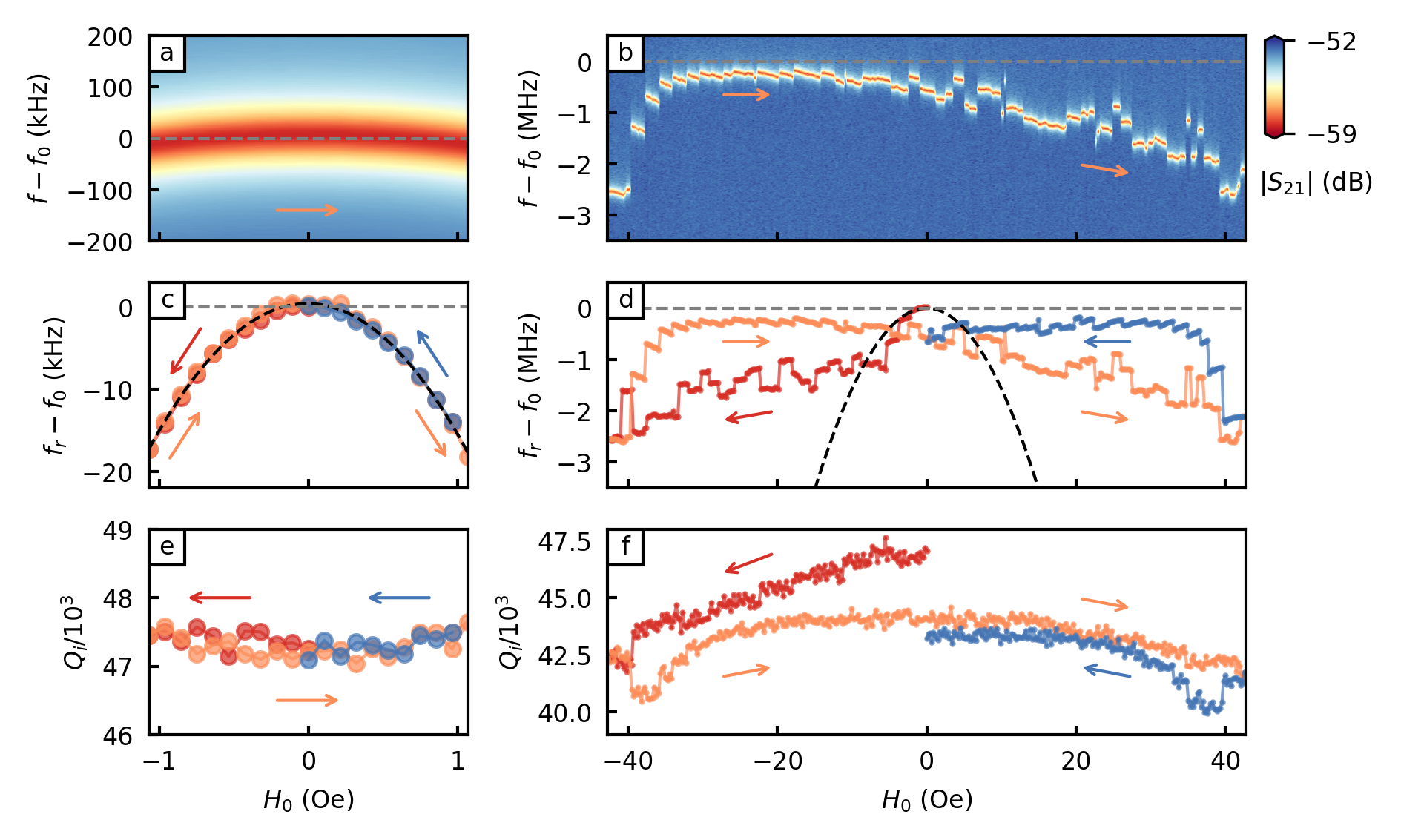}
\caption{{Magnetic-field dependence of resonator~\#3.} (a,b) Color maps of the transmission amplitude $|S_{21}|$ as a function of the {probe} frequency and the applied perpendicular field. The frequency axis is plotted relative to the zero-field resonance $f_0 = 3.832$~GHz {(horizontal gray dashed line).} {Resonance frequency} $f_r$ (c,d) and internal quality factor $Q_i$ (e,f)  extracted from line-shape fits during a full hysteresis cycle: sweeping from $0$ to $-H_{\mathrm{max}}$, then to $+H_{\mathrm{max}}$, and back to $0$. Arrows and colors indicate the sweep direction starting with the red curves. The color maps in (a,b) correspond to the orange lines shown in the subsequent panels. The black dashed curve is a parabolic fit to the low-field data in panel (c), extrapolated {to higher fields} in panel (d).}
\label{fig: From_H}
\end{center}
\end{figure*}

\subsection{Small magnetic fields: Meissner regime}

Figure~\ref{fig: From_H} {illustrates the evolution of the resonance peak in an external magnetic field applied perpendicular to the chip plane. At low fields---in our device,} for $|H_0|\lesssim 1$~Oe {(see Fig.~\ref{fig: From_H}~(a) and (c))---the resonant frequency decreases smoothly following an approximately parabolic law. This response reflects the suppression of the superconducting order parameter by Meissner screening currents and the associated increase of the penetration depth (kinetic inductance)}~\cite{Healey_2008}.
Within the magnetic field range considered, the internal quality factor shown in Fig.~\ref{fig: From_H}~(e) remains approximately constant, within an accuracy of about one percent estimated from the experimental data fit.
Importantly, within this low-field regime the resonance parameters return to their initial values once the field is removed.

\subsection{Large magnetic fields: Abrikosov vortices}

We extend the measurements {to perpendicular fields up to $40$~Oe. Before each hysteresis cycle, the sample is warmed above the critical temperature $T_c$ and then cooled to the base temperature ($18$~mK) in zero field, i.e., prepared in the zero-field-cooled (ZFC) state. As the field increases, the frequency deviates from the parabolic Meissner trend (Fig.~\ref{fig: From_H}(b) and (d)), and the first abrupt downward jump appears at $H_0\simeq -1.9$~Oe. We associate this event with the onset of vortex penetration.}

Using {a field step of $0.2$~Oe, we resolve that the dependence $f_r(H_0)$ consists of smooth segments separated by sharp jumps.} Up to $H_0\approx -10$~Oe, the resonance frequency remains close to the low-field parabolic extrapolation (black dashed line in Fig.~\ref{fig: From_H}(d)). 
Further increase in the magnetic field amplitude leads to frequency jumps in both directions. The overall averaged field dependence becomes linear. At a field of -40~Oe, the frequency significantly exceeds the parabolic extrapolation from the Meissner regime.
The quality factor shown in Fig.~\ref{fig: From_H}~(f) tends to decrease linearly as the field increases. The large error in determining this value makes it impossible to distinguish individual jumps.

After reaching $H_0\approx -40$~Oe, we reverse the sweep direction. The frequency then increases through a sequence of jumps and reaches a maximum before the field changes sign. The quality factor also increases when the field returns to zero.
Similar hysteretic behavior has been reported previously \cite{Bothner_2012, Bothner_2017} and is captured by the NBI (Norris–Brandt–Indenbom) critical-state model taking into account the inhomogeneous {distribution of microwave currents in a coplanar waveguide. In particular, immediately after field reversal vortices of opposite polarity can enter near the edges where the microwave current density is maximal, strongly affecting both $f_r$ and $Q_i$. Notably, after completing the cycle the resonance does not return to its initial zero-field value (gray dashed line in Fig.~\ref{fig: From_H}~b and d), indicating that a finite number of} vortices remains trapped.

\begin{figure*}[ht!]
\begin{center}
\includegraphics[width=16cm]{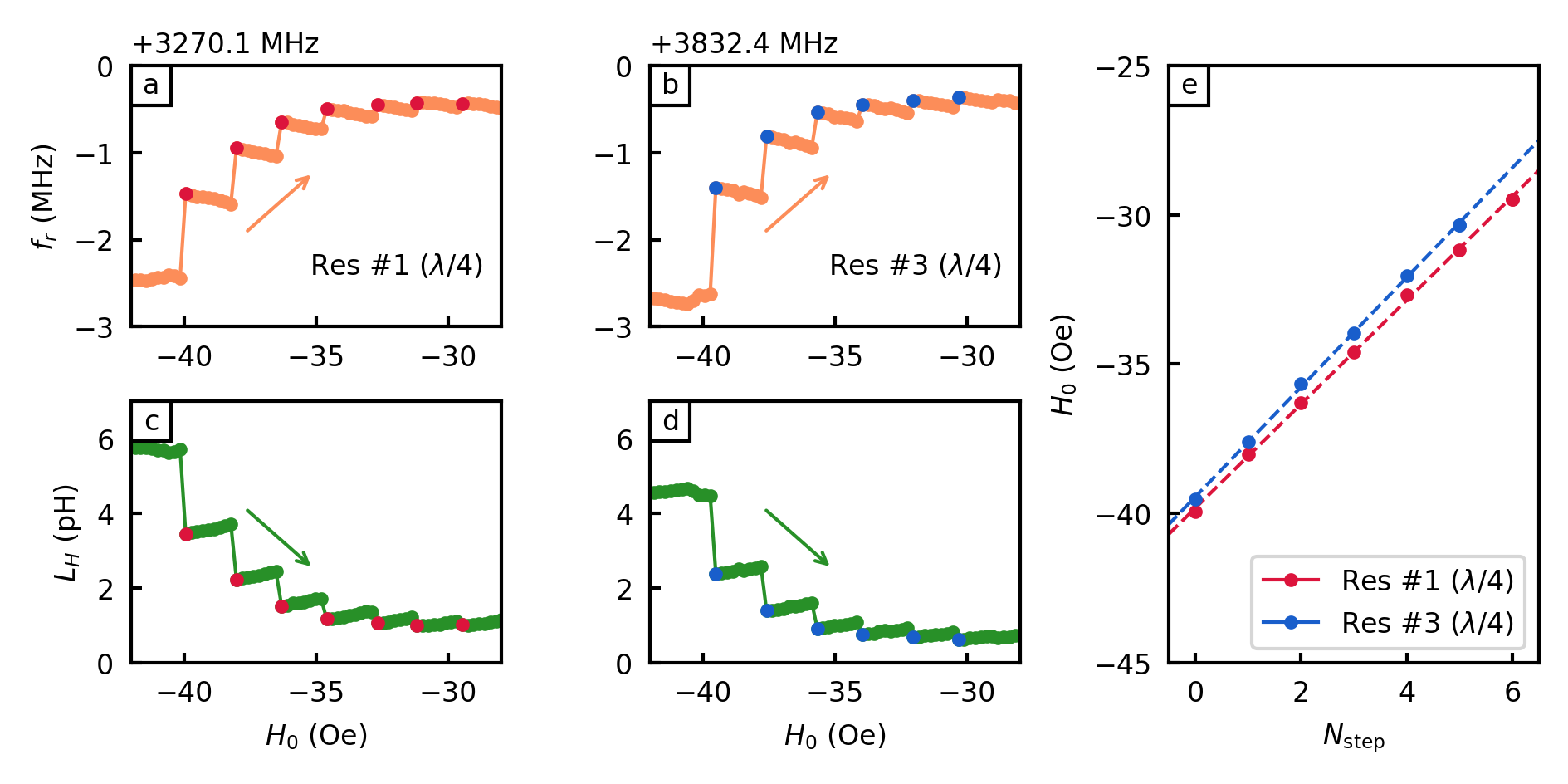}
\caption{Equidistant jumps in the resonance frequency. Panels (a) and (b) show the dependence of the resonant frequency on the magnetic field for the fundamental ($\lambda/4$) mode of resonators \#1 and \#3, respectively. Panels (c) and (d) present the behavior of the additional inductance as a function of magnetic field calculated from the frequency change. Arrows indicate the magnetic field sweep direction. The data in panel (b) correspond to a segment of the plot in Fig. \ref{fig: From_H}~(d). Color-coded markers denote data points following a sharp frequency jump. Panel (e) plots, using markers, the magnetic field value at which a frequency jump occurs versus the jump sequence number. Dashed lines represent linear fits to the experimental data.}
\label{fig: Periodic_steps}
\end{center}
\end{figure*}

We repeated the measurements with the microwave power increased and decreased by an order of magnitude and observed equivalent behavior of the resonator curve. This indicates that the microwave current does not contribute substantially to the vortex distribution compared to the external magnetic field. Furthermore, this implies that the vortex dynamics are primarily local, without a flux-flow regime.

\subsection{Equidistant frequency jumps}

We performed the field-dependent measurements for all resonators on the chip. The measurement procedure consists of successive frequency sweeps over all observable resonance dips at a fixed field, followed by a stepwise change of the field. Figure~\ref{fig: Periodic_steps}~(a) and (b) shows the results for two representative resonators (Res~\#1 and Res~\#3). To formalize the jump-detection procedure we use quantitative criteria: a jump was defined as a frequency change exceeding 33~kHz between neighboring measurement points, comparable to the resonance linewidth. Data analysis reveals that the positions of the jumps differ among different resonators, which indicates that the observed effect is not associated with a common central waveguide and instead points to a local origin of the phenomenon.

Nevertheless, the behavior of the resonant frequency remains similar for both resonators. To evaluate the effect of vortices, we calculated the change in the total inductance of the resonators $L_H$ due to the change of the magnetic field (see Fig.~\ref{fig: Periodic_steps}~(c) and (d)) based on the expression for frequency in Eq.~\eqref{eq: definition_fr_Qi}. For this calculation, we used the value of the capacitance per unit length $C = 0.16$~nF/m obtained from the resonator geometry, which we assume to remain constant. We found that the inductance change due to jumps diminishes monotonically and approaches zero as the field amplitude decreases, with the maximum value for both resonators being $\approx$~2~pH.

A notable result of these measurements is the observation of an approximately constant spacing between successive jumps along the magnetic-field axis, which we refer to as equidistance. This effect is reproducible over different field ranges ($H_{\text{max}} = 10, 20, 40$~Oe) and for all resonators, although it should be noted that a degree of stochasticity is present, occasionally leading to deviations from perfect equidistance. Such a non-ideal reproduction can be observed by comparing the orange and blue curves in Figure \ref{fig: From_H}~(d). Despite the symmetric path of the magnetic field (with the opposite sign), the jumps do not occur at the same positions, although the average behavior of the resonant frequency and the characteristic step size are preserved.

For quantitative analysis, Fig.~\ref{fig: Periodic_steps}(e) shows the magnetic-field value at which each jump occurs as a function of its ordinal number. A linear fit to this dependence yields a characteristic jump spacing of 1.7~Oe for resonator Res~\#1 and 1.8~Oe for resonator Res~\#3, indicating that the spacing is essentially the same for both resonators, especially given the 0.2~Oe step size. We note that the equidistant behavior emerges immediately after reversing the sweep direction of the magnetic field and persists approximately until the resonance frequency reaches its maximum value within the given field sweep cycle.

\begin{figure*}[ht!]
\begin{center}
\includegraphics[width=16cm]{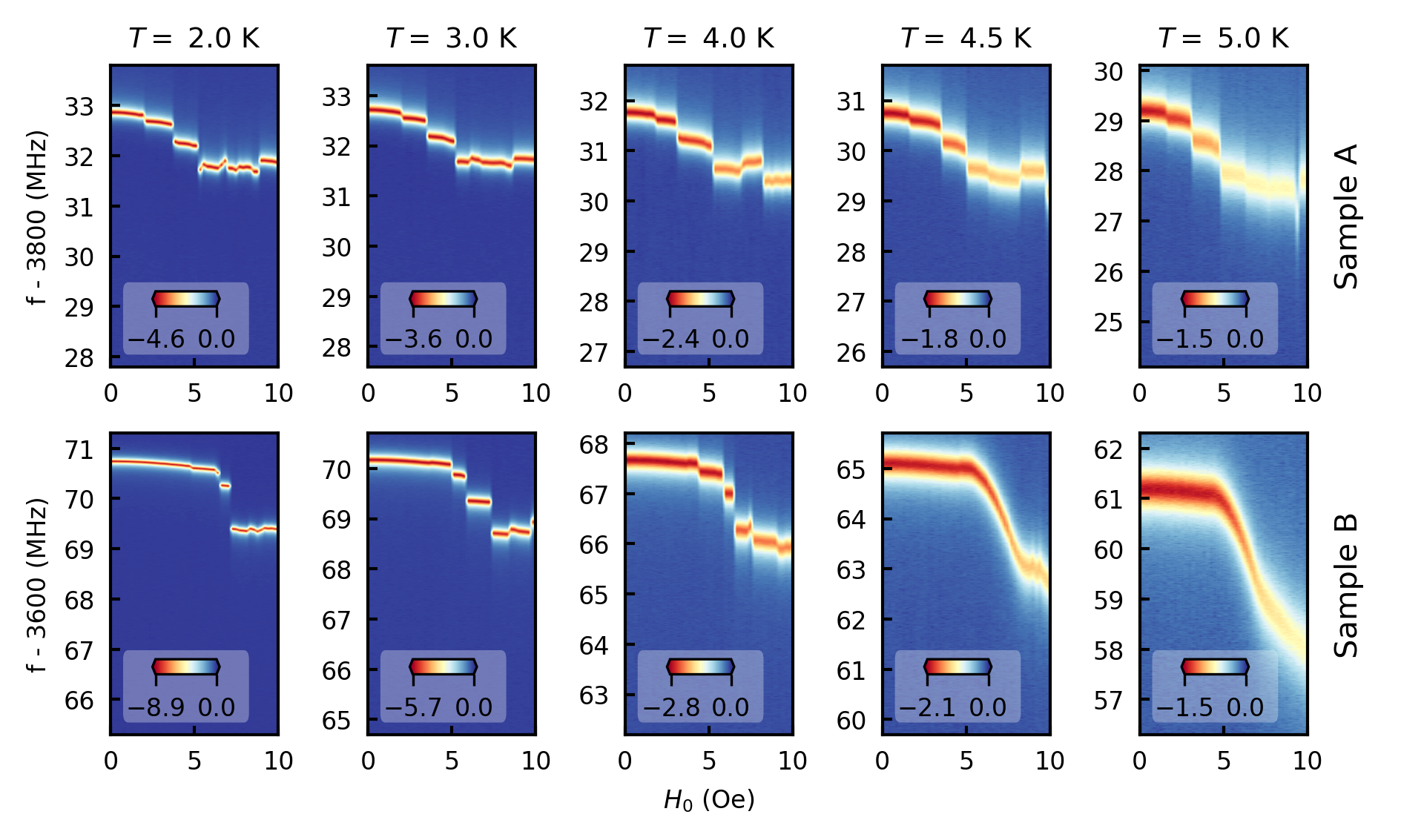}
\caption{Field-induced evolution of the resonance peak at different temperatures. Each panel shows $S_{21}$ (dB) as a function of probe frequency and the applied field swept from 0 to 10 Oe. The top row corresponds to sample~A, and the bottom row to sample~B; the resonators have identical geometries (Res \#3) but are fabricated on different substrates. All panels use the same axis ranges ($6$~MHz along the frequency axis and $10$~Oe along the field axis). The field is swept from $0$ toward positive values. Before each temperature point, the sample is warmed above $T_c$ to remove trapped flux and then cooled at zero field.}
\label{fig: FromVirginState}
\end{center}
\end{figure*}

\subsection{Temperature dependence of vortex entry}

We investigate the vortex-entry process as a function of temperature in both sample~A and sample~B. To avoid the influence of trapped vortices, each measurement begins from a zero-field-cooled (ZFC) state---warming above $T_c$ and then cooling in zero magnetic field. Results for two resonators of identical geometry, one fabricated on each sample, are summarized in Fig.~\ref{fig: FromVirginState}.

In sample~A, jump-like features persist up to at least $5$~K. With increasing temperature, the size of continuous sections between jumps decreases: for example, the first jump shifts from $H_0 \simeq 1.9$~Oe at $2$~K to $H_0 \simeq 1.3$~Oe at $5$~K. At the same time, the overall influence of the magnetic field grows stronger at higher temperature; at $H_0 = 10$~Oe, both the resonance-frequency shift and the relative change in the quality factor are significantly larger at $5$~K than at $2$~K.

Sample~B displays a qualitatively different behavior. The jumps are not equidistant---nor does an equidistant series appear after reversing the field-sweep direction---and by $T = 4.5$~K the magnetoresponse becomes largely smooth. Although the resonators on both samples share the same geometry (Res \#3), the chips are fabricated on different substrates: Si/SiO$_2$ for sample~A and high-resistivity Si for sample~B. 
As a result, the resonance frequency is lower for sample B ($f_r = 3.671$~GHz at 18 mK) than for sample A because the dielectric constant of Si is larger than that of SiO$_2$, while the quality factor is higher for sample B ($Q_i = 1.3\times10^5$ at 18 mK), since the crystalline silicon substrate introduces less energy dissipation compared to the amorphous oxide layer.

\section{Discussion}

\subsection{Surface barrier for vortex entry/exit}

Our measurements show that $f_r(H_0)$ (and, in general, $Q_i(H_0)$) consists of smooth segments separated by abrupt jumps. One may attribute these jumps to Abrikosov vortex entry or exit events. Such abrupt behavior may be connected with a surface barrier for vortex entry and exit known as the Bean–Livingston barrier \cite{BeanLivingston_1964}. 
Although vortices become energetically favorable above the first critical field $H_{\mathrm{c1}}$, the barrier delays their penetration to higher external fields. For a thin superconducting strip with thickness $d \lesssim \lambda$ (where $\lambda$ is the London penetration depth) and width $W \gg \Lambda$, where $\Lambda = \lambda^2 / d$ is the Pearl length, the first critical field is given by $H_{\mathrm{c1}} = \Phi_0 / (4\pi W \Lambda)\cdot \ln(2\Lambda / \xi)$, and the theoretical estimate for the field at which vortices overcome the barrier is $H_{\mathrm{s}} = \Phi_0 / (4\pi \xi \sqrt{W \Lambda})$ \cite{Maksimova1997}, where $\Phi_0$ is magnetic flux quantum and $\xi$ is the superconducting coherence length. 
Note that both fields decrease as the strip width $W$ increases, a behavior attributed to strong field focusing at its edges. Therefore, by reducing the width of the central resonator line, one can expand the vortex-free field region \cite{Bothner_2017}, which corresponds to a parabolic dependence of the resonance frequency.
For our parameters (standard for magnetron sputtered niobium films $\lambda = 100$ nm, $\xi = 10$ nm and dimensions $d=100$ nm, $W = 30$ $\mu$m) at low temperature, we obtain $H_{\mathrm{c1}} = 1.6$ Oe and $H_{\mathrm{s}} = 92$ Oe. 
The field at which the Meissner response ceases in our experiments is found to be only slightly above the estimated $H_{\mathrm{c1}}$ and much lower than the penetration field $H_{\mathrm{s}}$. This indicates that the surface barrier for vortex entry is suppressed in our samples, likely due to the granular structure of the niobium films and boundary roughness on the order of $100\,\mathrm{nm}$ (see Appendix~\ref{appendix: SEM}).    

Determining the exact locations and number of Abrikosov vortices entering the resonator remains an open question.
The work of Nulens et al. \cite{Nulens_2023} demonstrates that vortices enter the resonator line and the ground plane in large numbers via avalanches within a magnetic field range comparable to ours. More recently, Shulga et al. \cite{Shulga2025} also report discrete jumps in the resonant frequency $f_r(H_0)$ for a Nb resonator. They attribute these jumps to the penetration of individual Abrikosov vortices into the specially narrowed part of a superconducting strip forming the resonator, an interpretation supported by NV-center magnetometry that directly imaged discrete vortex entry events with increasing magnetic field. Given the similarity in the $f_r(H_0)$ response, the applicable magnetic field range, and the resonator material parameters to our own devices, it is plausible that we are indirectly observing similar vortex entry/exit processes in the Nb strip.
In our study all resonators on sample A, along with resonator \#3 on sample B, incorporated a geometric constriction (width 5 µm, length 70 µm; see Fig.~\ref{fig: Sample_B} caption). However, measurements of a resonator \#7 without a constriction on sample B discussed in Appendix~\ref{appendix: sample B} reveal that the jump-like behavior and its characteristic magnitude are nearly identical. We therefore conclude that the jumps are not exclusively related to the constriction region.

\subsection{The effect of Abrikosov vortices on the resonator parameters}

Figure~\ref{fig: From_H} demonstrates that the resonant frequency and quality factor do not change proportionally as was demonstrated in \cite{Song_2009}, where resonators in field cooled (FC) state were studied. In our case, upon decreasing the magnetic field amplitude after the sweep reversal, the frequency recovers almost completely, whereas the quality factor recovers only by half (see the orange curve in the negative-field region of Fig.~\ref{fig: From_H} (d) and (f)). This indicates that the active (dissipative) and reactive contributions of vortices to the resonator impedance change non-proportionally with the magnetic field.
We propose the following mechanism to explain this behavior. A vortex and an antivortex pinned in close proximity without annihilating mutually compensate their circulating currents, thereby significantly reducing their net contribution to the inductance. However, the dissipation, primarily associated with the vortex cores, remains additive, equivalent to the contribution of two individual vortices.
Consequently, we conclude that at the field reversal point in Fig.~\ref{fig: From_H} (at approximately -40 Oe), antivortices abruptly enter the resonator. This leads to a sharp increase in the resonant frequency and a simultaneous sharp decrease in the quality factor. As the field magnitude is further reduced, the antivortices begin to partially annihilate with the existing vortices, resulting in a recovery of the quality factor.

A precise theoretical calculation of the vortex contribution to the kinetic inductance and, consequently, to the resonance frequency shift is a complex task. It requires accounting for pinning on defects and vortex-vortex interactions. As an initial step, we estimated this contribution within a simplified framework: considering a narrow strip ($W \ll \Lambda$) without pinning (see Appendix \ref{Ap:model:vortex:Z}). In this case, the expression for the frequency shift due to $N$ non-interacting vortices can be derived from Eq. \eqref{eq:single:vort:res:freq:shift}:
\begin{equation}
\delta f \approx N \cdot \delta f^{v} \sim - \dfrac{N}{H_{0}}.
\end{equation}
While $N$ itself depends on the magnetic history and field, it remains constant as long as no vortices enter or exit the film. Under this condition, the frequency shift follows the dependence $\delta f \sim -H_0^{-1}$, which qualitatively describes the smooth decrease in frequency observed between the discrete jumps in the experiment (see Fig.~\ref{fig: Periodic_steps} (a,b)).

Further, we performed numerical calculations based on the Ginzburg-Landau equation (see Appendix \ref{Ap:numerical:vortex:Z}) under the same simplified assumptions. These calculations yield inductance values consistent with the analytical estimate. Furthermore, the modeling demonstrates that the presence of pinning on defects reduces the individual vortex contribution, as it restricts vortex motion under the influence of the current.

However, a correct interpretation of the experimental resonator frequency shift data requires consideration of key factors neglected in these initial estimates. Main among these is the highly inhomogeneous distribution of both the screening and the transport RF currents in the actual device geometry (see eq.~\eqref{eq:jH:wide} and \eqref{eq:j0:wide}). Nevertheless, to establish the relevant scale, we applied the expression \eqref{eq:Lkv_wide}. 
There \( H' \) corresponds to the field of the vortex exit, which generally depends on their distribution in the film. When the external field reaches \( H' \) the inductance increases infinitely. We note that in numerical calculations we also observe a strong increase in the inductance as \( H' \) is approached. However, in the experiment, $f_r(H)$ between jumps is nearly linear rather than hyperbolic. This suggests that at the moment of a jump,  $H'$ for this vortex configuration is at least about 10~Oe higher. We substitute this estimate into formula \ref{eq:Lkv_wide} and find that one vortex contributes an inductance on the order of 0.01~pH.
This value is two orders of magnitude smaller than the characteristic change in the total resonator inductance observed experimentally (see Fig.~\ref{fig: Periodic_steps}~c and d). From this comparison, we conclude that each observed jump in the resonance frequency corresponds to the entry of a large number of vortices into the sample ($\delta N \sim 10^{2}$).

\subsection{Origin of the equidistant jumps}

A notable finding of our research is that after reversing the field sweep, the jump fields form an almost equidistant sequence with $\Delta H\simeq 1.8$~Oe which is close to the first critical field.
We propose that each jump corresponds to reaching local magnetic field near the strip field $H_{\mathrm{local}}=H_{c1}$ leading to the entry of antivortices. In this picture, once a jump occurs the system relaxes to a new state with a field near the zero around the strip; the next jump is triggered when the shielding-current distribution, which changes approximately linearly with the applied field, again reaches a critical condition. 
A physically similar mechanism was responsible for the quasi-periodic avalanches of vortices \cite{Shvartzberg_2019} in the superconducting ring, when the difference between the applied field and the average field inside the central hole
reaches a threshold level. 
A similar edge-barrier instability was reported in thin films of PdBi\(_2\) and NbSe\(_2\), where slow field sweeps produced spikes in the attenuation and velocity of surface acoustic waves (SAWs) at intervals near \(H_{c1}\), attributed to abrupt vortex avalanches followed by relaxation \cite{cpl_42_10_100706}.
The threshold mechanism is also discussed in Ref.~\cite{Vodolazov_2003}, where the vorticity in the ring changes when the current density/supervelocity in the ring exceeds a critical value.

This mechanism of equidistant jumps assumes that, within a given field range, antivortices enter the resonator uniformly along its length, without forming avalanches. In our case, such behavior is consistent with the estimated low surface barrier. The formation of vortex avalanches is also known to depend on the magnetic field sweep rate \cite{jiang2023sensitivity}, which in our experiment was very low ($\sim 0.2\,\mathrm{Oe/s}$). Moreover, we did not observe any significant changes when varying the power of the microwave signal. Additionally, in Ref.~\cite{Nulens_2023}, where vortex penetration in coplanar resonators was investigated using magneto-optical imaging, no avalanches were observed during field sweeps from $H_{\max}$ down to zero within a certain field interval (down to a field denoted in that work as $H_{\mathrm{III}}$), although avalanches reappeared upon subsequent field increase.

At higher fields, the equidistant pattern breaks down, and irregular upward and downward frequency jumps emerge at positive fields (see the orange curve in Fig.~\ref{fig: From_H} (d)). We attribute this to deeper penetration of the antivortex front into the film, where vortex dynamics are governed by random pinning. In contrast, near $-H_{\max}$, variations in the number of vortices within the resonator are primarily controlled by the superconductor boundary. Due to the strongly nonuniform microwave current distribution, the response of a vortex depends sensitively on its distance from the boundary. Consequently, abrupt changes in the resonator characteristics may arise from sudden collective vortex motion (e.g., depinning) perpendicular to the boundary, even in the absence of new vortex entry. In addition, vortices of opposite polarity remain trapped in the film (see Appendix \ref{appendix: MFM} about our other study with similar niobium films). The observed upward frequency jumps with increasing field are therefore most likely associated with vortex--antivortex annihilation in the film interior.

\subsection{Role of film quality and vortex pinning: samples A vs B}

Figure~\ref{fig: FromVirginState} highlights a pronounced difference between the magnetic-field responses of samples~A and~B. We attribute this divergence primarily to differences in the superconducting film quality and the corresponding pinning landscape.

Vortex pinning in niobium films arises from multiple factors, including thickness modulations, grain boundaries, and nanoscale disorder. Recent scanning vortex microscopy has revealed that pinning centers in Nb films can form an extended correlated ``nano‐network'' rather than being randomly distributed \cite{Hovhannisyan_2025,larionov2025peculiarities,2025_Aladyshkin}. Due to pinning, after magnetic field sweeping, vortices are distributed non‐uniformly within the film. At the same time, the microwave current across the resonator cross‐section is also non‐uniform. Consequently, the contribution to the resonator impedance depends not only on the total number of vortices but also on their spatial distribution, which is governed by the local pinning strength. The granular structure of the film also weakens the surface barrier for vortex entry. Nevertheless, sample A exhibits robust and nearly equidistant steps, suggesting that the vortex-entry edge region is characterized by a comparatively uniform pinning landscape, allowing a reproducible threshold condition for successive penetration events.

In contrast, sample B shows a chaotic, irregular response. SEM images reveal a smaller grain size in sample B compared to sample A (Appendix~\ref{appendix: SEM}). Moreover, temperature‐dependent resistance measurements (Appendix~\ref{appendix: DC}) show that sample B has a lower superconducting transition temperature and a broader transition than sample A, indicative of enhanced disorder and inhomogeneity. The residual‐resistance ratio is also smaller for sample B, consistent with a higher defect density. Such microstructural differences are known to strongly affect vortex penetration and avalanche dynamics \cite{Oh_2024, APL10.1063/5.0282694_2025}. Together with a more gradual edge, this may explain the absence of equidistant steps in sample B. Furthermore, thermal activation (vortex creep) effectively weakens pinning and increases dissipative response at elevated temperatures \cite{Coffey_Clem_1991}. In the more disordered pinning potential of sample B, this effect is particularly pronounced and likely leads to the observed smearing of the steps at higher temperatures.

\section{Conclusion}
We have experimentally investigated the influence of a perpendicular magnetic field on superconducting coplanar resonators and observed a pronounced staircase-like magnetoresponse associated with Abrikosov vortex dynamics. Beyond the reversible Meissner regime, the entire resonance peak undergoes abrupt jumps as a function of the applied magnetic field. Upon reversal of the field sweep, these jumps form an almost equidistant series with a characteristic spacing $\Delta H \simeq 1.7$--$1.8$~Oe. While this spacing is observed for all studied resonators, the absolute fields at which the jumps occur differ between them, indicating the intrinsically local character of the underlying mechanism. 
These observations, supported by our theoretical estimations of the vortex inductance, point out to multiple-vortex entry and exit events.
A comparison of two chips fabricated from Nb films of different quality shows that the equidistant staircase is preserved over a broad temperature range in a higher-quality, more homogeneous film, whereas it is strongly degraded in a lower-quality, more inhomogeneous film, where the steps progressively smear out upon warming. 

The non-proportional and qualitatively different responses of the resonant frequency and the quality factor point to a complex interplay between vortex and antivortex configurations and the microwave properties of the resonator. Taken together, our results indicate that a comprehensive description of the observed behavior must simultaneously account for the discrete nature of individual vortex events and for the collective dynamics of a large number of vortices in the presence of pinning.

We expect that these findings will motivate further experimental and theoretical studies of vortex dynamics in planar superconducting structures. A predictive understanding of such complex vortex systems is essential for reliably assessing their impact on superconducting circuits and for enabling controlled and scalable use of vortices in emerging superconducting logic and computing architectures.

\acknowledgments

We thank Sergey Grebenchuk and Igor Golovchanskiy for their valuable advice and assistance in the early stages of this study.
This work was carried out using equipment from the MIPT Shared Facilities Center.
The research was supported by the Russian Science Foundation project No. 23-72-30004 (cryogenic measurements) and the Ministry of Science and Higher Education of the Russian Federation Project No. 075-15-2025-010 (analysis of experimental data, approximations)

\appendix

\section{Kinetic inductance of a single vortex: analytical estimations}\label{Ap:model:vortex:Z}

To estimate the changes in kinetic inductance and related shifts in resonance frequency we use the following model. We consider a superconducting narrow strip with length $l$, width $W$, and thickness $d$ ($l \gg W \gg d$ and $\lambda^{2} \gg W d$) in a perpendicular dc field $H_{0} \mathbf{e}_{z}$ and Abrikosov vortices in this strip. Within standard theoretical approaches \cite{Gittleman_1966,GorkovLP1975,KopninNB2001}, the vortex equation of motion can be written as (we neglect the inertial effects associated with the vortex mass, creep effect and intervortex interaction): 
\begin{equation}\label{eq:vort:mov}
    \eta \mathbf{v} = \dfrac{\Phi_{0}}{c} [ \mathbf{j}_{tr} \times \mathbf{e}_{z} ] - \nabla U_{\text{eff}} (\mathbf{R}),
\end{equation}
where $\eta$ is viscosity coefficient \cite{KopninNB2001}, $(\Phi_{0} / c) [ \mathbf{j}_{tr} \times \mathbf{e}_{z} ]$ is Lorentz force and $U_{\text{eff}}$ is an effective potential that includes both the interaction of vortex with pinning centers arising from sample inhomogeneity and the interaction of vortex with Meissner screening current and the sample edges, $\mathbf{R}$ and $\mathbf{v}$ denote the vortex position and velocity.

Assuming that the vortex undergoes small oscillations around its equilibrium position $\mathbf{R}_{0}$, we approximate the potential by a harmonic form $U \approx \eta \omega_{k} (\mathbf{R} - \mathbf{R}_{0})^{2} / 2$, where $\omega_{k}$ is the depinning constant and depends strongly on the pinning mechanism \cite{GolosovskyM1994}. With this approximation, Eq.~\eqref{eq:vort:mov} in the Fourier representation ($\partial_{t} \rightarrow i \omega)$ takes the form (the tilde indicates the Fourier component):
\begin{equation}
    (1 + \dfrac{\omega_{k}}{i \omega}) \eta \mathbf{\tilde{v}} = \mathbf{\tilde{F}}_{L}.
\end{equation}
Using expression for electric-field induced by a single vortex moving $\mathbf{\tilde{E}}_{v} = - \Phi_{0} [ \mathbf{\tilde{v}} \times \mathbf{e}_{z} ] / (l W)$ we obtain the vortex contribution to the impedance from $N$ vortices:
\begin{equation}\label{eq:vortices:Z}
    Z^{v}_{tot} = \sum\limits_{i = 1}^{N} Z^{v}_{i}, \qquad Z^{v}_{i} = \dfrac{\Phi_{0}^{2}}{c^{2} W^{2} d \eta} \dfrac{1}{1 + \dfrac{\omega_{k,i}}{i \omega}}.
\end{equation}
Eq.~\eqref{eq:vortices:Z} shows that for weak effective pinning, $\omega_{k,i}\ll \omega$, the impedance $Z_{i}^{v}$ is predominantly resistive while its reactive part behaves capacitively $\propto \omega^{-1}$. Whereas for strong effective pinning, $\omega_{k,i}\gg \omega$, $Z_{i}^{v}$ mainly contributes to the inductive response of the sample.

We next consider the case of strong pinning (purely inductive vortex response) and contribution to the characteristic pinning frequency $\omega_{k}$ that is common to all vortices and originates from the Meissner screening currents and from the vortex interaction with the sample edges. In the weak-screening approximation, the interaction energy between a vortex and the Meissner current $j_{s,M} = - c / (4 \pi \lambda^{2}) H_{0} x \mathbf{e}_{y}$ can be estimated as $U_{M} = \Phi_{0} H_{0} x^{2} / (8 \pi \lambda^{2})$. The edge contribution can be written as $- \nabla U_{\text{edges}} = \Phi_{0}^{2} / (16 \pi \lambda^{2} W) \tan (\pi x / W)$, $|x| \leq W / 2$ \cite{ShmidtVV1974,Maksimova1998}.

Thus, for a vortex oscillating near the strip center ($|x|\ll W$) we obtain
\begin{equation} 
\omega_{k} \approx \dfrac{1}{\eta} \dfrac{\Phi_{0}}{4 \pi \lambda^{2}} (H_{0} - H'), \qquad H' = \dfrac{\pi \Phi_{0}}{4 W^{2}}.
\end{equation}
The corresponding single-vortex contribution to the kinetic inductance and to the resonance-frequency shift reads
\begin{equation} 
\label{eq: L_K^v}
L_{k}^{v} = L_{k0} \frac{ \Phi_{0} }{(H_{0} - H') l W}, \qquad L_{k0} = \dfrac{4 \pi \lambda^{2} l}{W d},
\end{equation} 
and
\begin{equation}\label{eq:single:vort:res:freq:shift}
\dfrac{\delta f^{v}}{f_{0}} = - \dfrac{1}{2} \left(1 + \frac{W d}{\lambda^2}\right)^{-1} \dfrac{\Phi_0}{(H_{0} - H') l W} . 
\end{equation}

The divergence as $H_0\to H'$ reflects the rapid increase of the vortex displacement at a fixed ac current amplitude (note, that it is consequence of made approach $|x|\ll W$).

For a wide strip $W d\gg \lambda^{2}$, the edge force does not admit a simple analytical form and for simplicity we neglect it. Using the screening and transport current densities \cite{Kupriyanov1975,Clem1998}
\begin{align}
\label{eq:jH:wide}
j_{s,M}=-\frac{cH_0x}{2\pi d\sqrt{(W/2)^2-x^2}}, 
\\
\label{eq:j0:wide}
j_{tr}=\frac{I}{\pi d\sqrt{(W/2)^2-x^2}},
\end{align}

one finds in the quasistationary limit for the vortex located near the center of the strip
\begin{equation}
\label{eq:Lkv_wide}
L_k^v=\frac{2\Phi_0}{W(H_0 - H')}
\end{equation}
where we introduce the vortex exit field as $H'$. 
Above expression is not applicable at very low value of $|H_0 - H'|$ when the vortex exits the strip and one may expect large increase of $L_k^v$ at finite magnetic field.

\section{Kinetic inductance of a single vortex: numerical calculations}\label{Ap:numerical:vortex:Z}

\begin{figure}[t]
\includegraphics[width=0.48\textwidth]{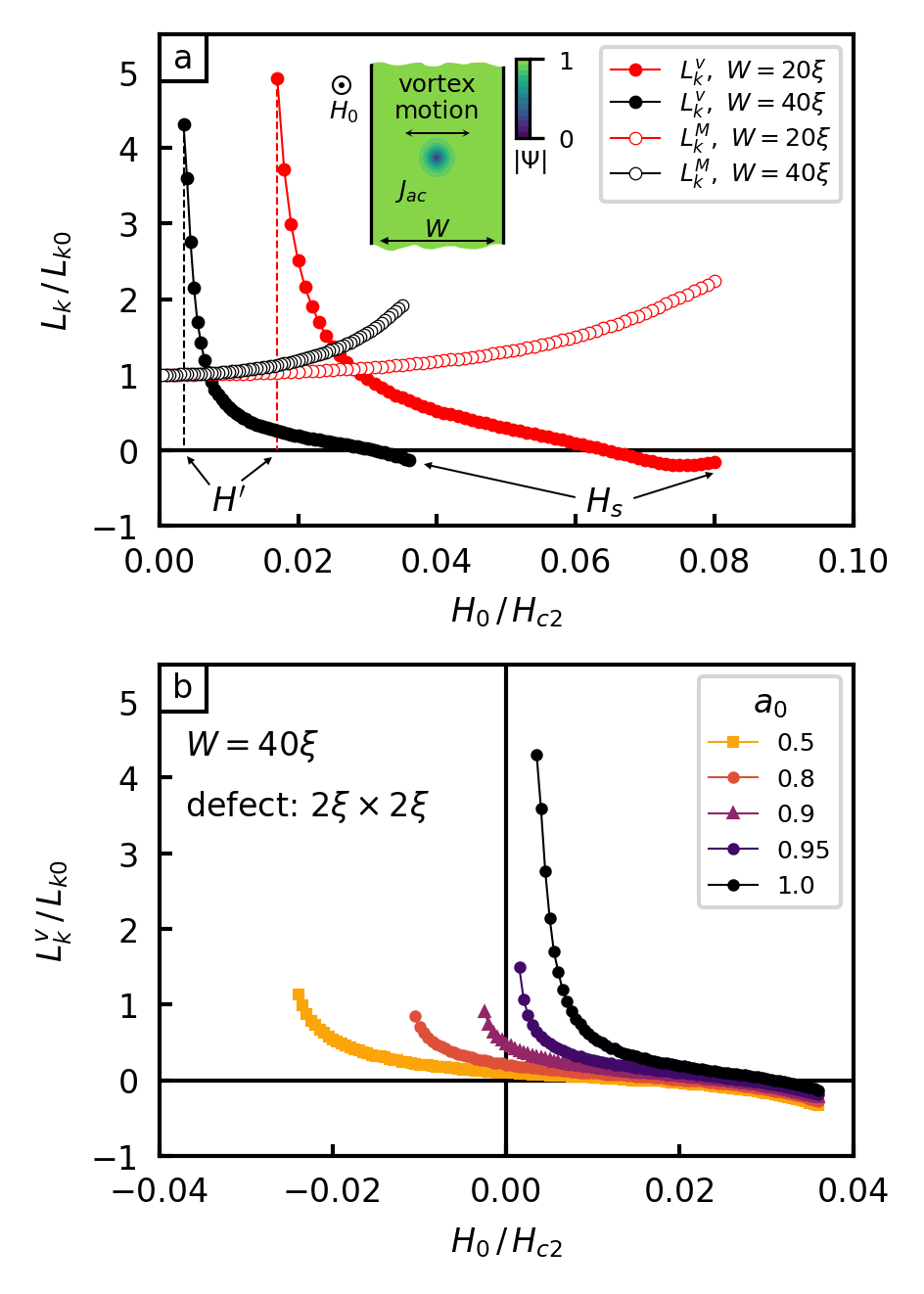}
\caption{(a) Field-dependent kinetic inductance of a vortex ($L_k^v$, solid symbols) and of a narrow strip without a vortex ($L_k^M$, open symbols) obtained using the Ginzburg-Landau approach. In the inset we show sketch of the strip with a vortex. (b) Field-dependent kinetic inductance of a vortex pinned in the center of the superconducting strip for different pinning strengths, controlled by the parameter $a$ in the Eq.~\eqref{eq:TDGL:Psi}.}
\label{fig: vortex_Lk}
\end{figure}

\begin{figure*}[ht!]
\begin{center}
\includegraphics[width=16cm]{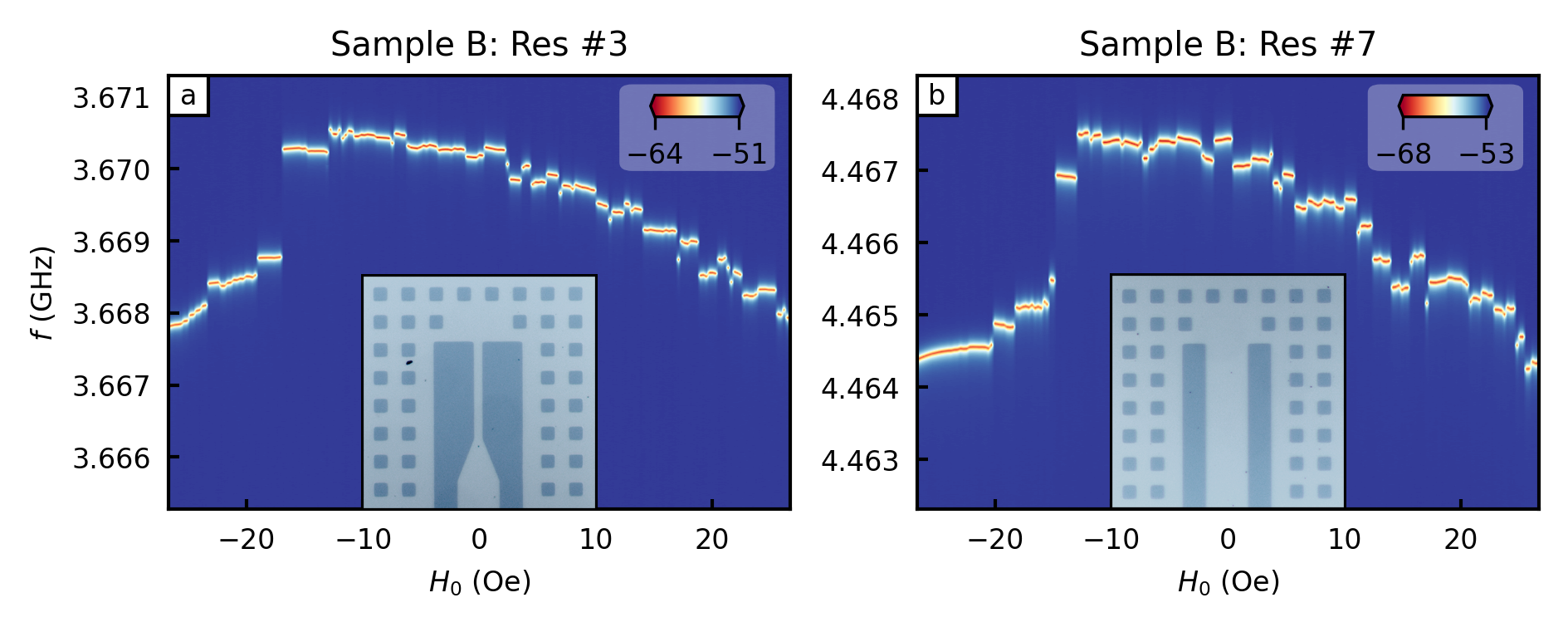}
\caption{Color maps of the transmission amplitude, $|S_{21}|$ (dB), as a function of signal frequency and applied magnetic field for Res~\#3 (a) and Res~\#7 (b) of the sample B. The inserts show images of the resonators region where the central line connects to the ground. All resonators have the central line width $W = 30~\mu$m and the gap to the ground $G = 17~\mu$m. Resonator \#3 contains a narrowed section in the central line with width $5~\mu$m and length $70~\mu$m.}
\label{fig: Sample_B}
\end{center}
\end{figure*}

In the quasistationary limit the kinetic inductance of the vortex can also be found using time-dependent Ginzburg-Landau equation for the superconducting order parameter $\Psi$ \cite{KopninNB2001}:
\begin{equation}\label{eq:TDGL:Psi}
\left(\frac {\partial }{\partial t} +i\varphi \right)\Psi
=(\nabla + i \mathbf{A})^2 \Psi +(a(\mathbf{r})-|\Psi|^2)\Psi.
\end{equation}
In Eq.~\eqref{eq:TDGL:Psi} $\Psi$ is normalized in units of
$\Psi_0=3.07k_BT_c(1-T/T_c)^{1/2}$, distance is in units of
temperature-dependent coherence length $\xi$, vector potential is
in units of $\Phi_0/2\pi \xi$. In these units the magnetic field
is scaled by $H_{c2}=\Phi_0/2\pi \xi^2$ and the current density by
$j_0=c\Phi_0/8\pi^2\lambda^2\xi$. Time is scaled in units of the
Ginzburg-Landau relaxation time $\tau_{GL}=2\hbar k_BT_c/\pi\Psi_0^2$,
the electrostatic potential $\varphi$, is in units of
$\varphi_0=\hbar/2e\tau_{GL}$. In our calculations we use the
superconductor-vacuum boundary conditions at the edges of the
strip: $(\nabla - i {\bf A})\Psi|_n=0$, $\nabla
\varphi|_n=0$ and superconductor-normal metal boundary conditions at
its ends: $\Psi=0$, $\nabla \varphi|_n= -I/Wd$.
Because we consider narrow strip limit ($\lambda^2/d > W$) we neglect
contribution to ${\bf A}$ from screening and transport current and
choose ${\bf A}=(0,H_0x,0)$. To find $\varphi$ we solve the following equation
\begin{equation}\label{eq:TDGL:phi} 
\Delta \varphi = {\rm div}\left({\rm Im}(\Psi^*(\nabla+{i}{\bf A})\Psi)\right),
\end{equation}
which is coming from the conservation of the total current in the strip, i.e. ${\rm div} {\bf j}=0$.

In the numerical calculations, we vary the width of the strip, $W$, and set its length as $l=5W$. We also consider the case where a pinning center with size $2\xi \times 2\xi$ is located at the center of the strip. Inside the pinning area, we choose $a({\mathbf{r}})=a_0<1$, while outside, $a({\mathbf{r}})=1$ in Eq.~\eqref{eq:TDGL:Psi}. Physically, this corresponds to a variation in the local critical temperature.

To find the kinetic inductance
\begin{equation}
L_k(H_0)=\frac{c^2}{I}\frac{\partial F_s(H_0,I)}{\partial I},
\end{equation}

we calculate current-dependent superconducting free energy
\begin{equation}
F_s(I,H_0)=-d\frac{H_{cm}^2}{8\pi}\int |\Psi|^4 dxdy
\end{equation}
where we integrate over the width and length of the strip
($H_{cm}=\Phi_0/2\sqrt{2}\pi \xi \lambda$ is thermodynamic
magnetic field) and $|\Psi|(x,y)$ corresponds to stationary solution of Eqs.~(\ref{eq:TDGL:Psi}, \ref{eq:TDGL:phi}) at fixed $H_0$ and $I$. To find $L_k^v$ we calculate $L_k(H_0)$ for the strip without vortex ($L_k^M$ - Meissner state), with vortex and subtracted one from the other. 

In Fig.~\ref{fig: vortex_Lk}, we plot our results for strips with different widths, where kinetic inductance is scaled in units of $L_{k0}$ - see Eq. \eqref{eq: L_K^v} (for our parameters $L_{k0}\sim 0.13$ pH). The dependence $L_k^{M}(H_0)$ is nonlinear in the Meissner state due to the depairing effect of screening currents and it is doubled at relatively large  $H \sim H_s$. Here, $H_s$ is the magnetic field corresponding to the first vortex entry; in the London model, $H_s=\Phi_0/ 2\pi \xi W$ \cite{Maksimova1998}. At $H\sim H'$, there is a rapid increase of $L_k^v$, which has the same reason as in the analytical model. When $H<H'$, vortex exits the strip, and $L_k^v=0$.

At a field around $H_s$ $L_k^v$ becomes negative. We attribute this effect to the partial compensation of screening currents by currents flowing around the vortex, which reduces nonlinear effects. This leads to a smaller overall value of the sum $L_k^M+L_k^v$ compared to $L_k^M$ alone. When $H>H_s$, vortices enter the strip, and our calculations are terminated.

From Fig.~\ref{fig: vortex_Lk}(b), it follows that the pinning center does not significantly change $L_k^v$, at least in our model. The value of $L_k^v$ remains approximately equal to $L_{k0}$ 
and increases as $H_0$ decreases. The pinning center reduces the maximum value of $L_k^v$ because vortex displacement is now restricted by the pinning region, and allows the vortex to exist even at zero or negative magnetic fields.

Using the above results, we can estimate the value of the single vortex kinetic inductance in our system. Since $d \sim \lambda \sim 100$~nm, we have $\Lambda \sim 100$ nm $ \ll 30~\mu$m. In Eq.~\eqref{eq:Lkv_wide} $H_0$ has a meaning of the magnetic field outside the strip. In our case magnetic field outside the strip does not coincide with external magnetic field due to screening by ground plate and pinned vortices. For estimation we take $H_0=1-10$~Oe (it is around of $H_{c1}$ for our strip) and find $L_k^v \sim 0.13-0.013$~pH. 

In the experiment, the inductance varies by $0.1-1$~pH (see Fig. \ref{fig: Periodic_steps}). If position of the pinning center is located close to the edge of the strip it may lead to increase of $L_k^v$ due to locally larger $j_0$ (see Eq.~\eqref{eq:j0:wide}). In any case we assume that the variation of inductance in our experiment could be related to the exit of many vortices from the strip or entry of many antivortices, which compensate fully (if they annihilate vortices) or partially (if they are pinned on some distance from vortices) contribution of vortices to the inductance.

\begin{figure}[t!]
\begin{center}
\includegraphics[width=8cm]{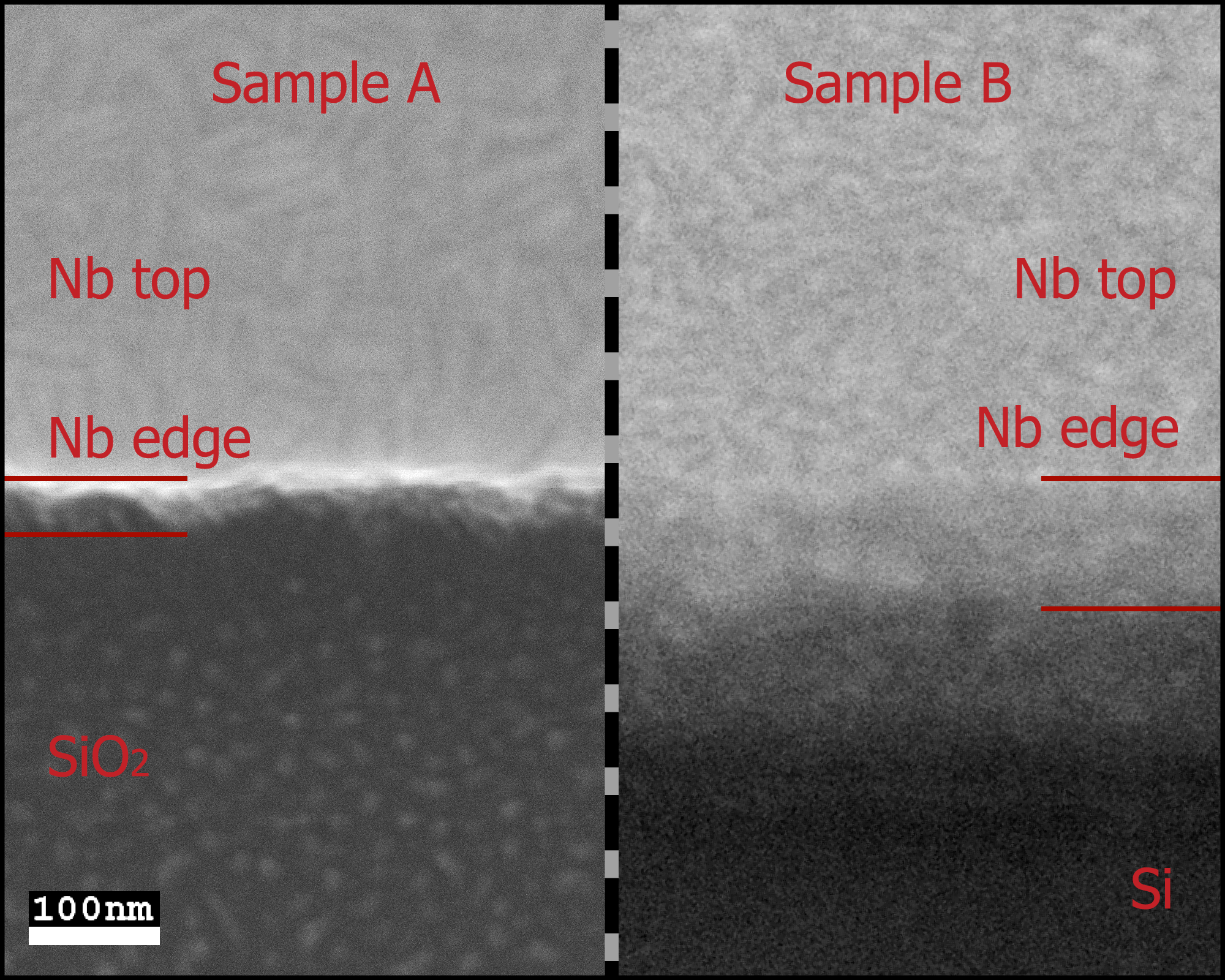}
\caption{Scanning electron microscope (SEM) images of samples A and B. The images show the boundary between the central resonator line and the etched gap. Both images were acquired in a top-down geometry. The manually drawn lines indicate the edges of the superconducting film.}
\label{fig: SEM}
\end{center}
\end{figure}

\section{Methods}
\label{appendix: Methods}
Both chips were fabricated from $100$~nm thick niobium films deposited by magnetron sputtering. Sample~A was grown on a Si/SiO${}_2$ substrate, whereas sample~B was grown on high-resistivity Si. The coplanar structures were patterned by optical lithography and plasma etching through a polymer mask. We used a CF${}_4$ + O${}_2$ plasma for sample~A and an SF${}_6$ + C${}_2$H${}_2$ plasma for sample~B. The lower film quality of sample~B may originate from a degraded vacuum or contamination of the Nb source during deposition.

The measurements were performed in a dilution refrigerator with a base temperature of $18$~mK. A perpendicular magnetic field was generated by a superconducting solenoid with a conversion factor of $1.068$~Oe/mA, driven by a Yokogawa GS200 dc current source. The transmission parameter $S_{21}$ was measured using a vector network analyzer connected to the sample holder via coaxial microwave lines. The input line included $40$~dB of cryogenic attenuation. Unless stated otherwise, the measurements were performed at a readout power of $-30$~dBm~($1~\mu$W).
With all the powers used in the experiment, the resonators were measured in multiphoton mode $n_{\text{ph}} \gg1$ \cite{McRaeCRH2020ThePerformance}.

\section{Magnetic field measurements on Sample B}
\label{appendix: sample B}

A resonator with identical geometry to the one on sample A (Res \#3) was fabricated on sample B (also labeled as Res \#3). Additionally, a resonator without a constriction at the junction between the center line and the ground plane was also produced (Res \#7). Figure~\ref{fig: Sample_B} shows the dependence of the resonator response on the magnetic field for a field sweep sequence: from $-H_{\text{max}}$ to $+H_{\text{max}}$.

\section{Scanning electron microscope images}
\label{appendix: SEM}

\begin{figure}[t!]
\begin{center}
\includegraphics[width=6cm]{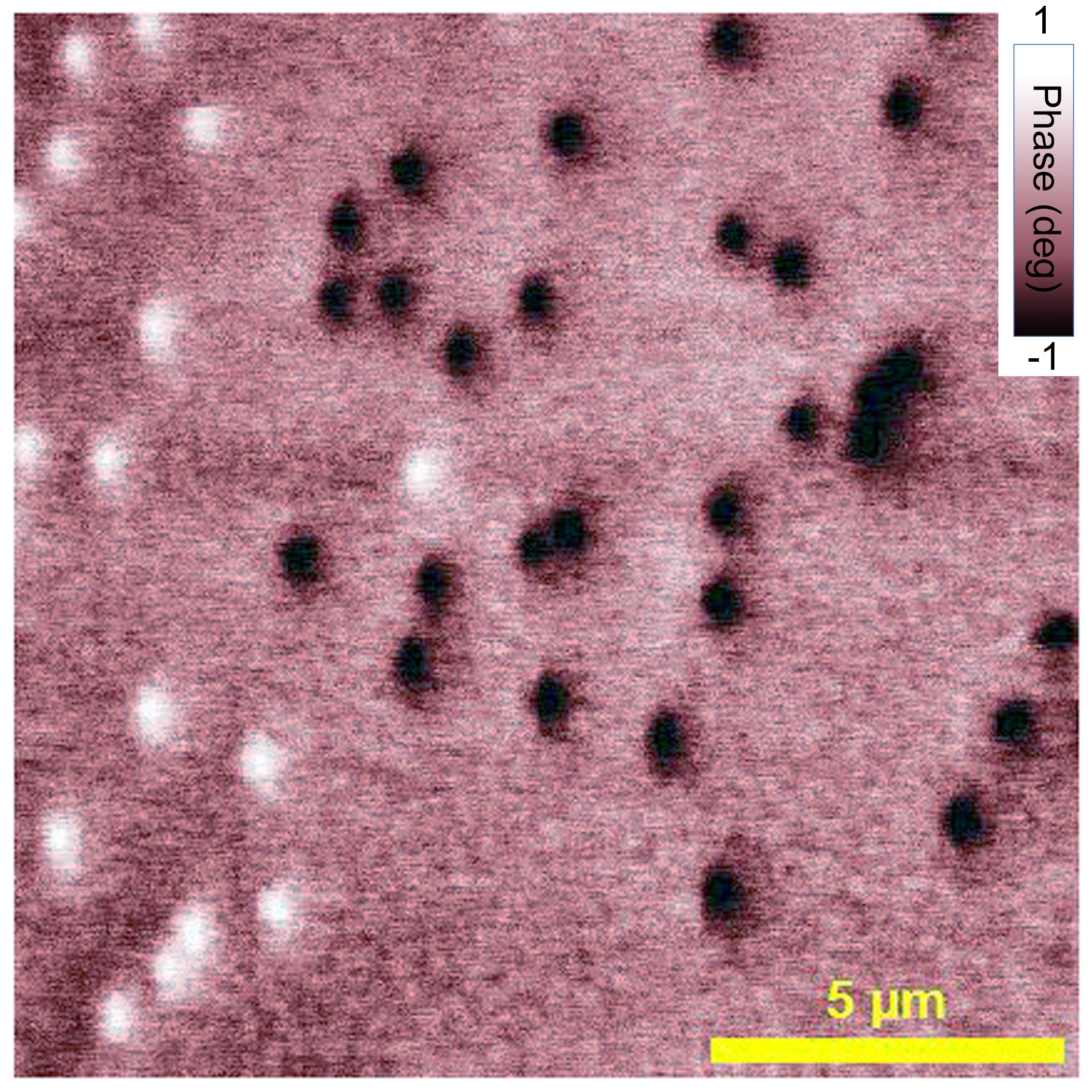}
\caption{A 15\textmu m$\times$15\textmu m magnetic force microscopy image showing local redistribution of the pinned vortices and antivortices on the surface of a 100~nm thick niobium film. The image was obtained at a temperature of 4.2 K and an external magnetic field of -25~Oe with a prehistory after the introduction of +100~Oe. White circles correspond to vortices with a positive field direction, and black --- with a negative field.}
\label{fig: MFM}
\end{center}
\end{figure}

Scanning electron microscope (SEM) images of both samples are presented in Figure~\ref{fig: SEM}.
On the niobium surface, elongated grain-like features are clearly resolved. For sample A, the lateral grain dimensions can be estimated to be approximately $100 \times 10\,\mathrm{nm}$. In contrast, sample B exhibits a finer grain structure forming a more densely packed network.

Although the images were recorded in a nominally top-down (plan-view) geometry, the boundaries of the superconducting region remain clearly visible. This indicates that the edges are not perfectly vertical but instead exhibit a finite sidewall slope. Such an effect can arise from gradual erosion of the resist mask during the etching process. Indeed, different resist materials and plasma chemistries were used during the fabrication of the two samples.
From a geometrical analysis of the SEM images, the edge inclination angle is estimated to be approximately $75^\circ$ for sample A and $45^\circ$ for sample B (where $90^\circ$ corresponds to an ideal vertical sidewall).

In both samples, the etching process was intentionally overextended in time. As a result, after complete removal of the niobium film, the plasma continues to etch the underlying substrate for a short period. In the figure, the manually drawn lines indicate the approximate extent of the boundary region corresponding to niobium, as inferred from contrast differences.

\section{Example of vortex/antivortex distribution}
\label{appendix: MFM}
In another study, we examined the surface of a 100~nm thick niobium film fabricated using the same magnetron sputtering method. Figure~\ref{fig: MFM} shows a magnetic field map obtained by magnetic force microscopy at a temperature of 4.2~K. This image primarily illustrates the possibility of vortices and antivortices coexisting in close proximity during film demagnetization without annihilation, which is hindered by pinning. In this image, the minimum distance between a vortex and an antivortex is 1.5~\textmu m. The measured film had dimensions of $5 \times 5$~mm, and the scanning was performed near its center.

\begin{figure}[t!]
\begin{center}
\includegraphics[width=8cm]{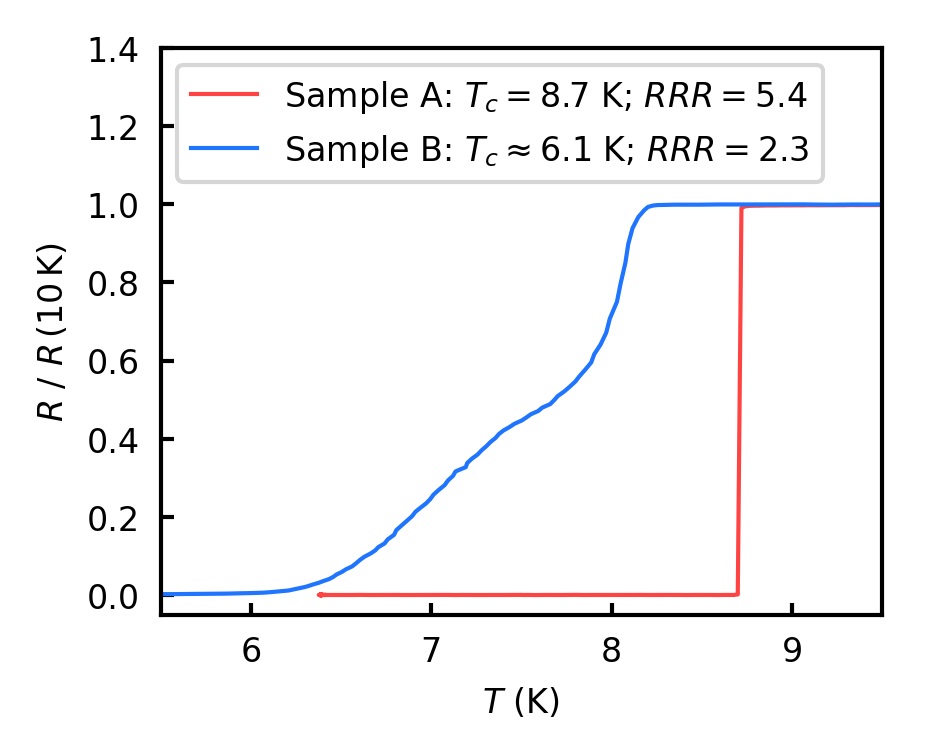}
\caption{Normalized resistance $R(T)/R(10\,\mathrm{K})$ for samples~A and~B at low temperatures illustrating the superconducting transition temperatures. The legend also shows residual-resistance ratios (RRR).}
\label{fig: R(T)}
\end{center}
\end{figure}

\section{R(T) for samples A and B}
\label{appendix: DC}

The temperature-dependent resistance of samples A and B was measured using a direct current method to determine their critical temperature, $T_c$, and the residual resistivity ratio, $RRR = R(300~\text{K}) / R(10~\text{K})$. As can be seen from the data shown in Figure \verb|\ref{fig: R(T)}|, sample A exhibits a sharp superconducting transition with a critical temperature of $T_c = 8.7~\text{K}$. In sample B, the resistance begins to decrease at approximately 8.1 K but drops to zero only at 6.1 K, which we define as its critical temperature, $T_c$.

\bibliography{References}

\end{document}